\documentclass[runningheads]{llncs}
\usepackage{comment}
\usepackage{graphicx}
\usepackage{booktabs}
\usepackage{multirow}
\usepackage{makecell}
\usepackage{subfigure}
\usepackage{amsmath}
\usepackage{marvosym}
\begin{document}
\title{Blind Surveillance Image Quality Assessment via Deep Neural Network Combined with the Visual Saliency}

\def\CICAISubNumber{100}  

\titlerunning{Blind Surveillance IQA via DNN Combined with the Visual Saliency}
%
\author{Wei Lu \and Wei Sun \and Wenhan Zhu \and Xiongkuo Min \and Zicheng Zhang  \and \\ Tao Wang \and Guangtao Zhai}
\authorrunning{W. Lu et al.}
%
\institute{Institute of Image Communication and Network Engineering,\\ Shanghai Jiao Tong University, China}
\maketitle              

\begin{abstract}
The intelligent video surveillance system (IVSS) can automatically analyze the content of the surveillance image (SI) and reduce the burden of the manual labour. However, the SIs may suffer quality degradations in the procedure of acquisition, compression, and transmission, which makes IVSS hard to understand the content of SIs. In this paper, we first conduct an example experiment (i.e. the face detection task) to demonstrate that the quality of the SIs has a crucial impact on the performance of the IVSS, and then propose a saliency-based deep neural network for the blind quality assessment of the SIs, which helps IVSS to filter the low-quality SIs and improve the detection and recognition performance. Specifically, we first compute the saliency map of the SI to select the most salient local region since the salient regions usually contain rich semantic information for machine vision and thus have a great impact on the overall quality of the SIs. Next, the convolutional neural network (CNN) is adopted to extract quality-aware features for the whole image and local region, which are then mapped into the global and local quality scores through the fully connected (FC) network respectively. Finally, the overall quality score is computed as the weighted sum of the global and local quality scores. Experimental results on the SI quality database (SIQD) show that the proposed method outperforms all compared state-of-the-art BIQA methods.

\keywords{Surveillance image  \and blind quality assessment \and deep neural network \and visual saliency.}
\end{abstract}

\section{Introduction}
\label{sec:intro}

With the rapid development of computer vision technology and the growing demand for security, recent years have witnessed the increasing popularity of the intelligent video surveillance system (IVSS), which can automatically analyze and understand the content of surveillance videos and thus greatly reduce the burden of the manual labour. Specifically, IVSS involves analyzing the videos using algorithms that detect, track, and recognize objects of interest \cite{sreenu2019intelligent}. However, the surveillance images (SIs) may suffer different degrees of degradations in quality due to the artifacts introduced to the SI acquisition and transmission process \cite{zhu2018siqd,leszczuk2012quality,jiang2018scalable,zhang2016best}, such as poor lighting conditions, low compression bit rates, etc. The SIs with low quality may lead to the poor performance of the computer vision tasks and thus make the IVSS difficult to effectively analyze the content of surveillance videos \cite{muller2005performance,aqqa2019understanding}. Hence, it is necessary to develop an effective quality assessment tool for SIs, which can help the IVSS to filter the low-quality SIs and improve the performance of the IVSS.

\begin{figure}[t]  
\centering
\subfigure[]{
\begin{minipage}[b]{0.475\textwidth}
\includegraphics[width=1\textwidth]{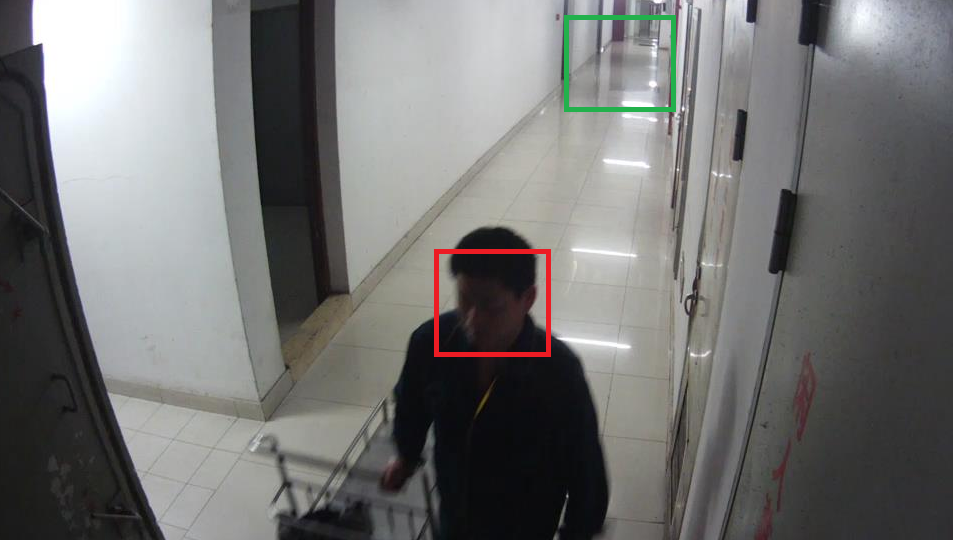}
\end{minipage}
}
\subfigure[]{
\begin{minipage}[b]{0.475\textwidth}
\includegraphics[width=1\textwidth]{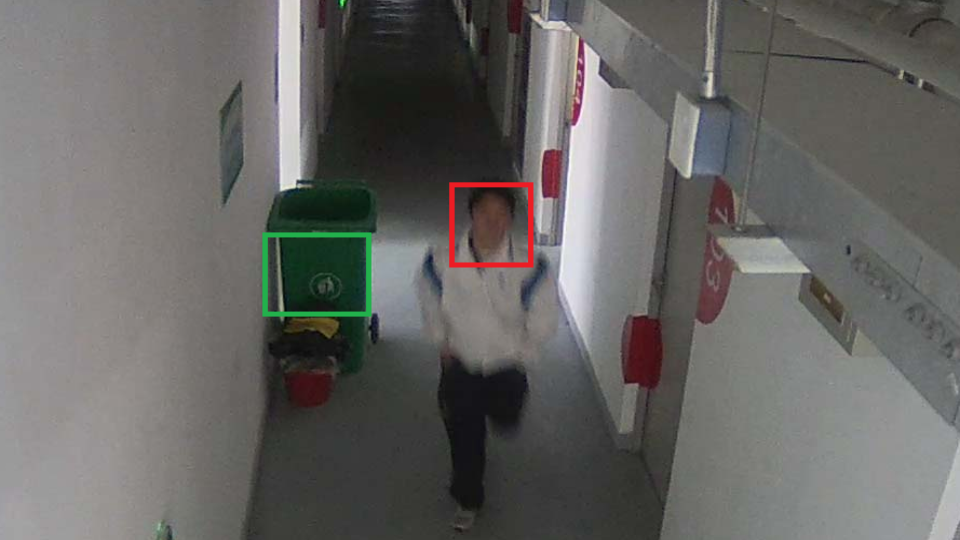}
\end{minipage}
}
\caption{Examples of the SIs with low image quality, which are from the SI quality database (SIQD) \cite{zhu2018siqd}. The red rectangle region is the blurred face and the green rectangle region is the clear background.} \label{fig1.1}
\end{figure}

Objective image quality assessment (IQA), which aims to automatically predict the perceptual visual quality of images, is a hot topic in the field of image processing \cite{sun2021blind}. Since the reference image is not available in the surveillance system, only blind IQA (BIQA) is suitable for qualifying the quality of SIs. Existing BIQA methods can be divided into two categories, which are hand-crafted feature based and deep learning based algorithms \cite{zhai2021perceptual}. Hand-crafted feature based BIQA models generally design quality-related features like natural scene statistics (NSS) features \cite{mittal2012no,fang2014no,liu2019unsupervised}, free energy features \cite{gu2014using,zhai2011psychovisual}, textures \cite{min2018blind,gu2014efficient}, etc., and then map these features into quality scores via a regression model such as support vector regression. Mittal $et\ al$. \cite{mittal2012no} utilized the statistical property in the spatial domain to predict the quality. Gu $et\ al$. \cite{gu2014using} proposed a BIQA metric using the free-energy-based brain theory and classical human visual system (HVS)-inspired features.  Deep learning based models usually adopt the convolutional neural network (CNN) to extract quality-aware features automatically and then regress them to quality scores with a fully connected (FC) network \cite{sun2021deep,wang2021multi,lu2022cnn}. Sun $et\ al$. \cite{sun2021deep} proposed a staircase structure to hierarchically integrate the features from intermediate layers into the final quality feature representation.

However, most existing BIQA models are developed for evaluating the images with synthetic distortions such as JPEG compression, Gaussian noise, etc., and do not perform well on the SIs with authentic and diverse distortions. Moreover, the distortions in the salient regions or regions of interest (ROI) have a greater impact on the overall visual quality of the SIs than that in the inconspicuous regions, when the human visual system (HVS) or IVSS tries to extract useful information from the SIs. As depicted in Fig.~\ref{fig1.1}, the SIs with a clear background and a blurred face generally have poor quality since the human and IVSS cannot recognize the very blur face. However, most existing BIQA methods do not consider this characteristic and have an inferior correlation with the quality of the SIs.

In this paper, we demonstrate that the quality of the SIs has a crucial impact on the performance of the intelligent video surveillance system through an example experiment (the face detection task) on the SI quality database (SIQD) \cite{zhu2018siqd}, and propose a visual saliency based deep neural network for the blind quality assessment of the SIs. To be more specific, we first compute the saliency map of each SI and select the local region with the maximum saliency weight. Second, the CNN is adopted to extract the quality-aware features for the whole image and the most salient local region, which are then mapped into the global quality and local quality score via the FC network respectively. Finally, the overall quality score of the SI is obtained by the weighted sum of global and local quality scores. Experimental results demonstrate that the proposed method outperforms state-of-the-art BIQA methods on the SIQD, which proves the effectiveness of the proposed method.

\begin{figure}[t]
\centering
\includegraphics[width=0.98\linewidth]{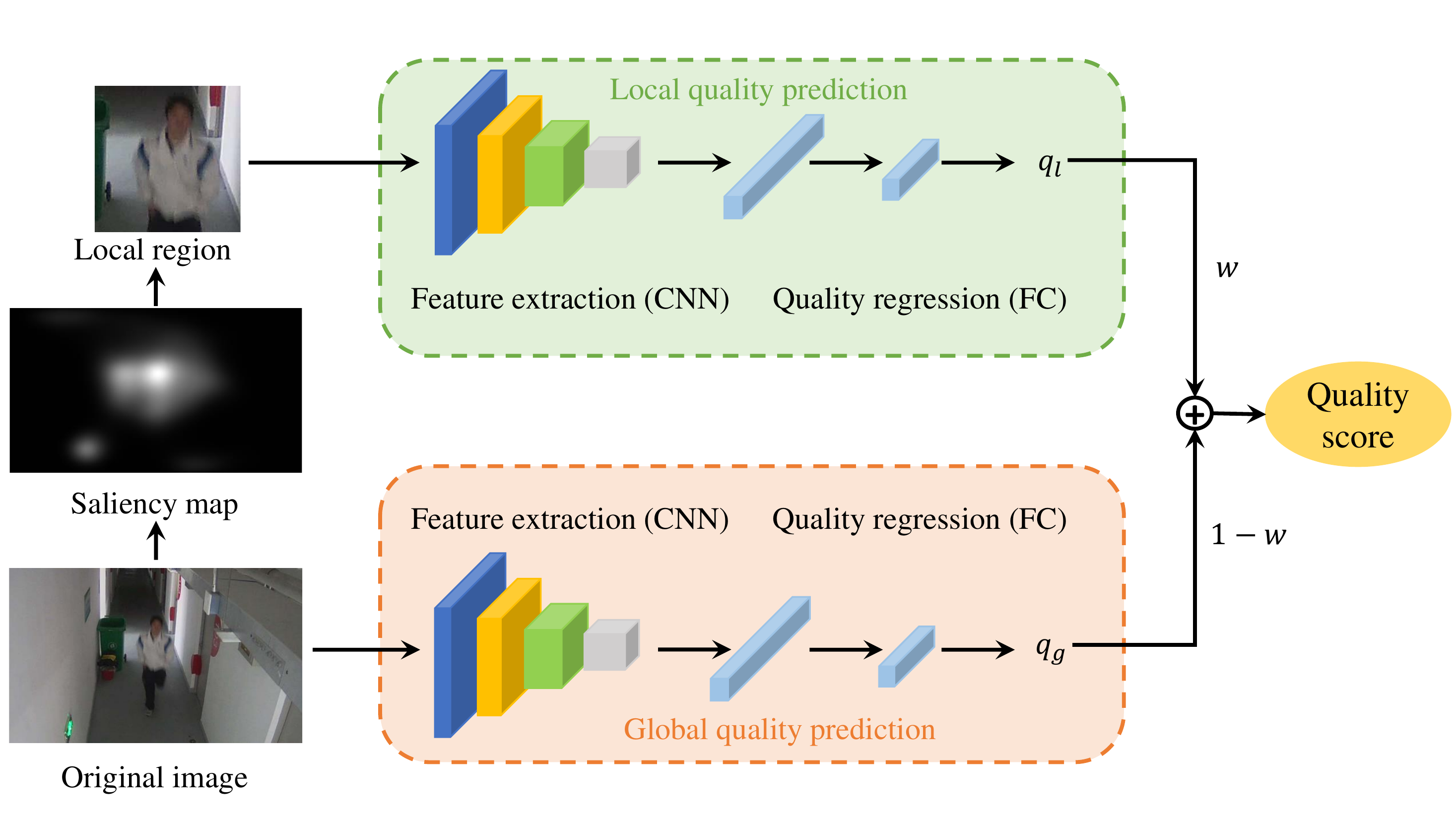}
\caption{The framework of the proposed method for the blind quality assessment of the SIs. }
\label{fig2.1}
\end{figure}

\section{Proposed Method}
\label{sec:method}

In this section, the proposed blind surveillance IQA model is described in detail. As shown in Fig.~\ref{fig2.1}, the framework of the proposed method mainly includes three stages: selecting the local region via the saliency map, extracting the quality-aware features for the whole image and local region, and predicting the overall quality score.

\subsection{Saliency-based Local Region Selection}

According to the characteristics of the HVS, different regions of the image obtain different degrees of attention. The salient regions have a great impact on the overall quality of the SIs since the salient regions usually contain rich semantic information, which is very important for machine vision algorithms. Hence, we propose to simultaneously predict the global quality of the whole image and the local quality of the salient region. 

As for the local region selection, we first compute the saliency map $S$ for the image $I$ using a state-of-the-art saliency model SimpleNet \cite{reddy2020tidying}, which is pretrained on a large saliency benchmark dataset SALICON \cite{jiang2015salicon}. The saliency value $S(i,j)$ of the pixel $(i,j)$ ranges in $[0,1]$, and the higher saliency value means it is more salient. Then we crop multiple image regions with the resolution of $224\times224$ through scanning the image $I$ both horizontally and vertically with a stride size of 32 pixels. The overall saliency weight of each image region $P$ is calculated as follow:
\begin{equation}
    sw = \sum_{x} \sum_{y} S(x, y), (x, y)  \in P.
\end{equation}

Finally, the image region with the maximum saliency weight $sw$ is selected for the local quality prediction.

\subsection{Quality-aware Feature Extraction}
\label{fe}

In recent years, CNN has demonstrated its powerful ability to solve various visual signal problems. Recent BIQA models mostly adopt the deep learning based architecture, which uses a CNN backbone to extract quality-aware features of distorted images, and then uses a fully connected network to aggregate them to quality scores. The architecture could be trained in an end-to-end manner and has become dominant in BIQA field. Compared with the handcrafted features, the features extracted by CNN contain more semantic information, which are suitable for quality assessment of SIs since a high quality of the SI means we can get exact semantic information from it. 

In the proposed method, we adopt the ResNet \cite{he2016deep}, which is a commonly used backbone in existing BIQA models, to extract quality-aware features for the whole image and local region. Specifically, the ResNet18 pretrained on the ImageNet \cite{deng2009imagenet} is selected for the quality-aware feature extraction. First, we feed the whole image $I_{g}$ or local region $I_{l}$ into the CNN model and then obtain the feature map from the last convolutional layer:
\begin{equation}
\begin{gathered}
    F_{g} = {\rm CNN}_{global}(I_{g}), \\
    F_{l} = {\rm CNN}_{local}(I_{l}),
\end{gathered}
\end{equation}
where ${\rm CNN}_{global}$, ${\rm CNN}_{local}$ refer to the CNN models for global and local feature extraction respectively and they do not share the same parameters. $F_{g}$ and $F_{l}$ are respectively the feature maps extracted from the whole image and local region. Second, the feature maps are converted into the feature vectors through the spatial
global average pooling: 
\begin{equation}
\begin{gathered}
    f_{g} = {\rm GAP}(F_{g}), \\f_{l} = {\rm GAP}(F_{l}),
\end{gathered}
\end{equation}
where GAP(·) means the the spatial global average pooling, and $f_{g}$, $f_{l}$ refer to the extracted feature vectors for the whole image $I_{g}$ and local region $I_{l}$ respectively.

\subsection{Quality prediction}

With the extracted quality-aware features, we adopt two FC layers as the regression model to predict the image quality. Specifically, the features are mapped into the quality scores through two FC layers consisting of 128 and 1 neurons respectively:
\begin{equation}
\begin{gathered}
    q_{g} = {\rm FC}_{global}(f_{g}), \\q_{l} = {\rm FC}_{local}(f_{l}),
\end{gathered}
\end{equation}
where ${\rm FC}_{global}$ and ${\rm FC}_{local}$ are respectively the global quality regression model and local quality regression model. $q_{g}$ means the global quality and $q_{l}$ means the local quality. The overall quality score $q$ is computed as the weighted sum of the global and local scores:
\begin{equation}
    q = w \cdot q_{l} + (1-w) \cdot q_{g},
\end{equation}
where the weight $w$ for the local quality ranges from 0 to 1. Finally, we can train the feature extraction network and image quality regressor in an end-to-end training manner, and the euclidean distance is used as the loss function, which can be computed as:
\begin{equation}
    L=\left\|q-q_{label}\right\|^{2},
\end{equation}
where $q_{label}$ refer to the ground-truth quality score. 

\section{Experiments}
\label{experiment}

In this section, we first briefly introduce the latest SI quality assessment database SIQD \cite{zhu2018siqd}. Next, we test a face detection algorithm on different quality of SIs selected from the SIQD, and the experimental results show that the visual quality of the SIs has a crucial impact on the face detection performance. Then we compare the proposed method and other state-of-the-art (SOTA) BIQA methods on the SIQD database. Finally, we analyze how the weight of the local quality affects the final predictive performance.

\subsection{Test database}

The SIQD database \cite{zhu2018siqd} includes 500 in-the-wild SIs with different degrees of quality, which cover various scenarios and diverse illumination conditions. The objects in the SIs contain human, vehicle (license number), etc., and the database contains four different resolutions: 1920×1080, 1280×720, 704×576, and 352×288. The mean opinion scores (MOSs) of each SI range from 1 to 5, where the larger MOSs, the higher quality.

\subsection{The Effect of Visual Quality on IVSS}

Although there is a common consensus that the low visual quality images will lead to the performance drop of the IVSS, it is necessary to conduct the experiment to validate and quantify this impact. In the experiment, we take the face detection task as an example to investigate this impact. We first select 286 SIs which contain one or more faces from the SIQD database and annotate face regions in images with bounding boxes. Then, the pretrained MTCNN \cite{zhang2016joint} network is used to detect faces for these SIs with the ground-truth bounding boxes. Finally, we compare the face detection performance on the SIs of different levels of quality. Specifically, the bounding box detected by the MTCNN network is considered to be correct if and only if its Intersection over Union (IoU) with the ground truth bounding box is more than 50\%, and three common evaluation metrics are used to measure the face detection performance, which are Precision, Recall, and Accuracy.

\begin{table}[t]
\renewcommand\arraystretch{1.15}
\setlength\tabcolsep{6pt} 
\normalsize
\centering
\caption{The performance of the face detection on the SIs of different levels of quality in the SIQD database.}\label{tab3.1}
\begin{tabular}{c|ccc}
\toprule%
MOS & Precision & Recall & Accuracy \\
\hline
[1.0, 2.5] & 0.8203 & 0.3477 & 0.3231 \\
(2.5, 3.5] & 0.7778 & 0.5147 & 0.4487 \\
(3.5, 5.0] & 0.7200 & 0.9403 & 0.6885 \\
\bottomrule
\end{tabular}
\end{table}

The performance results on the SIs of different levels of quality are summarized in Table~\ref{tab3.1}. From Table~\ref{tab3.1}, in terms of the Recall and Accuracy metric, we first observe that the performance of high-quality SIs is significantly higher than that of low-quality SIs, which indicates that the face detection performance of the SIs is largely affected by their visual quality. Hence, it is necessary to use the BIQA metric to filter the low-quality SIs, which can help to improve the performance of face detection. In addition, we can see that the Precision value drops slightly as the visual quality increases, which may be because the misjudgments of the face detection model are more likely to occur in the background of high-quality SIs than that of low-quality SIs.

\subsection{Performance Comparison with SOTA BIQA Methods}
\label{performance}

\subsubsection{Evaluation Metrics}

We adopt Pearson Linear Correlation Coefficient (PLCC), Spearman Rank Order Correlation Coefficient (SROCC),  and Root Mean Squared Error (RMSE) to evaluate different BIQA methods. SROCC represents the prediction monotonicity, while PLCC and RMSE indicate the prediction accuracy. An excellent model should obtain values of SROCC and PLCC close to 1, and the value of RMSE near 0. Before calculating the PLCC and RMSE, the nonlinear four-parameter logistic function in \cite{seshadrinathan2010study} is applied to map the scores predicted by objective BIQA methods to the subjective scores.

\subsubsection{Experiment Setup}

As mentioned in Section~\ref{fe}, the ResNet-18 is used as the backbone for feature extraction. 
The proposed model is trained and tested on a server with Intel Xeon Silver 4210R CPU @ 2.40 GHz,
128 GB RAM, and NVIDIA GTX 3090 GPU. The proposed model is implemented in PyTorch. The Adam optimizer with the initial learning rate 0.0001 is used to train the proposed model. The epochs are set at 50 and the batch size is set at 8. As for the global quality prediction, we resize the resolution of the minimum dimension of
images as 480 while maintaining their aspect ratios, and crop the images at the resolution of 448×448 at the center. The weight parameter $w$ for the local quality is set as 0.2.

The proposed method is validated on the SIQD database. In the experiments, the SIQD database is divided into the training set of 80\% SIs and the test set of 20\% SIs. To ensure complete separation of training and testing contents, we assign the SIs belonging to the same scene to the same set. We randomly split the database for
10 times, and the average value of the above evaluation metrics
is computed as the final result.

\subsubsection{Compared Methods}

We compare the proposed method with eleven popular BIQA models, including:
\begin{itemize}
    \item Hand-crafted feature based BIQA models: BLIINDS-II \cite{saad2012blind}, BRISQUE \cite{mittal2012no}, CORNIA \cite{ye2012unsupervised}, DIIVINE \cite{moorthy2011blind}, GMLF \cite{xue2014blind}, HOSA \cite{xu2016blind}, SISBLIM \cite{xu2016blind}, and NFERM \cite{gu2014using}.
    
    \item Deep learning based BIQA models: SFA \cite{li2018has}, DBCNN \cite{zhang2018blind}, and HyperIQA \cite{su2020blindly}.
\end{itemize}
All compared methods are retrained on the SIQD database.

\subsubsection{Results}

\begin{table}[t]
\renewcommand\arraystretch{1.15}
\setlength\tabcolsep{6pt} 
\normalsize
\caption{Performance comparison of the proposed model and eleven BIQA methods on the SIQD database.}\label{tab3.2}
\centering
\begin{tabular}{c|c|ccc}
\toprule
Type & Methods  & PLCC &SROCC & RMSE \\
\hline
\multirow{8}*{\makecell[c]{Hand-crafted\\ based}}          & BLIINDS-II & 0.2059 & 0.1584 & 0.903  \\
    ~                           & BRISQUE    & 0.3256 & 0.3051 & 0.8726 \\
     ~                          & CORNIA     & 0.5641 & 0.5476 & 0.7619 \\
    ~                           & DIIVINE    & 0.2178 & 0.0223 & 0.9007 \\
    ~                           & GMLF       & 0.2058 & 0.0740  & 0.9030\\
    ~                           & HOSA       & 0.3273 & 0.2871 & 0.8720  \\
    ~                           & SISBLIM    & 0.5488 & 0.4206 & 0.7714 \\
    ~                           & NFERM      & 0.3925 & 0.2576 & 0.8488 \\
\hline
\multirow{5}*{\makecell[c]{Deep learning\\ based}} & SFA        & 0.8741 & 0.8702 & 0.4153 \\
    ~                           & DBCNN      & 0.8785 & 0.8727 & 0.4033 \\
    ~                           & HyperIQA   & 0.8687 & 0.8631 & 0.4478 \\
~ & Proposed   & \bfseries0.9025 & \bfseries0.8973 & \bfseries0.3927 \\ 
\bottomrule
\end{tabular}
\end{table}

\begin{figure}[!t]
\centering
\includegraphics[width=0.65\textwidth]{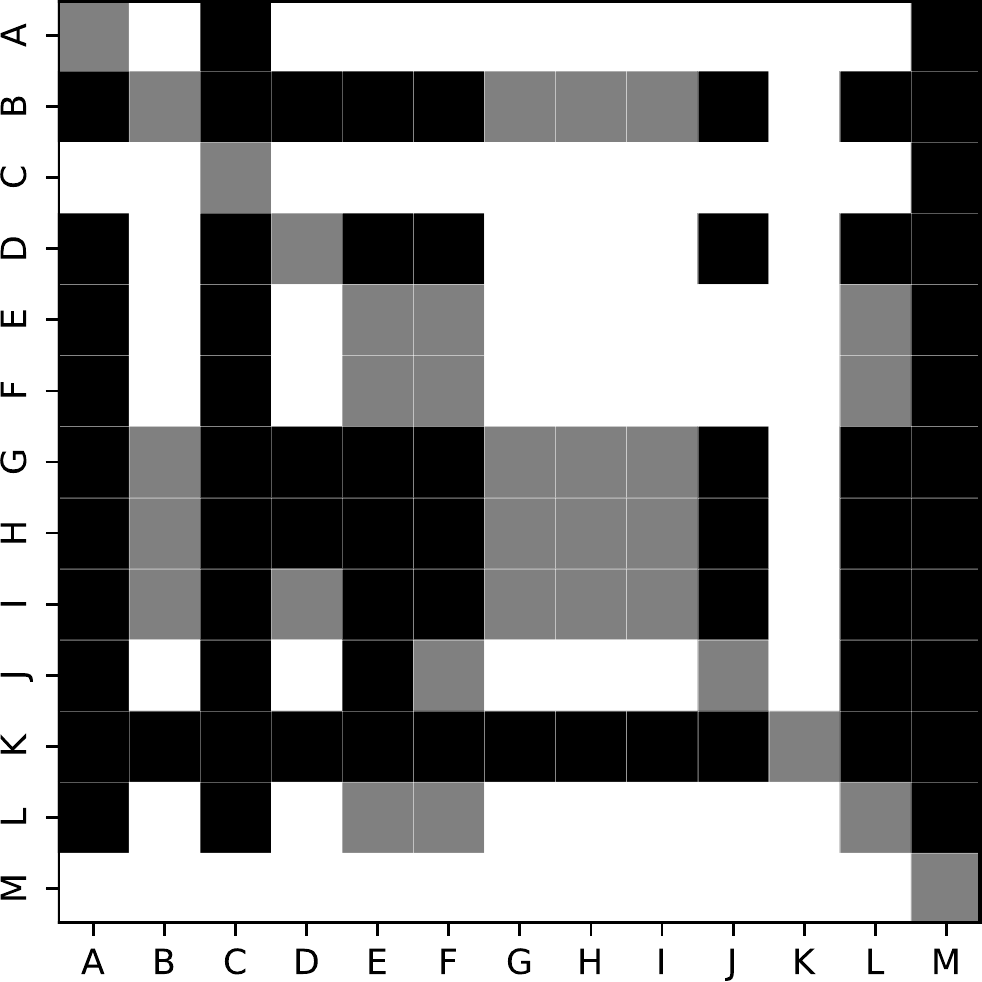}
\caption{Statistical significance comparison between the proposed model and
other BIQA methods on the SIQD. A black/white block $(i,j)$ means
the method at row $i$ is statistically worse/better than the one at column $j$. A gray block $(m,n)$ means the method at row $m$ and the method at $n$ are statistically
indistinguishable. The metrics denoted by A-M are of the same order as the compared metrics in Tabel~\ref{tab3.2}}
\label{fig3.1}
\end{figure}

The performance of the proposed method and compared BIQA models are depicted in Table~\ref{tab3.2}. We can observe that the proposed method outperforms all compared BIQA models, which demonstrates the effectiveness of the proposed model in the quality prediction of the SIs. Besides, the deep learning based methods perform better than the hand-crafted feature based methods and lead by a large margin, which indicates that the features extracted by CNN are more effective and more suitable for diverse distortions in the SIs. In addition, the proposed method performs better than compared deep learning based methods which do not consider the visual saliency. It can be concluded that the introduction of the visual saliency characteristics does improve the quality predictive performance of the SIs.

To further analyze the performance of the proposed method and other BIQA models, we conduct the statistical significance test in \cite{sheikh2006statistical} to measure the difference between the predicted quality scores and the subjective ratings. Fig.~\ref{fig3.1} presents the results of the statistical significance test for the proposed method and other BIQA models on the SIQD. We can clearly observe that the performance of our proposed BIQA model is statistically superior to other compared BIQA models on the SIQD.

\subsubsection{Ablation Study: the Importance of the Salient Region}

In this section, we mainly analyze how the relative weighting between the global quality and local quality affects the final predictive performance of the proposed model. Except for the weight $w$ for the local quality, the remaining experimental settings are set the same as that in Section~\ref{performance}. 

The experimental results are listed in Table~\ref{tab3.3}. From Table~\ref{tab3.3}, we can first see that the proposed method achieves the best performance when the weight $w$ is set as 0.2. Next, the proposed model without the global quality ($w=1$) performs only slightly worse than that without the local quality ($w=0$), and both of them perform worse than the proposed model with the global and local quality, which on the one hand indicates that the quality of the salient region has a great effect on the overall quality of the SIs and on the other hand indicates that it is necessary to simultaneously consider the global and local quality.

\begin{table}[t]
\renewcommand\arraystretch{1.15}
\setlength\tabcolsep{6pt} 
\normalsize
\caption{Performance comparison of different weights for the local quality.}\label{tab3.3}
\centering
\begin{tabular}{c|ccc}
\toprule
$w$  & PLCC &SROCC & RMSE \\
\hline
0   & 0.8768 & 0.8699 & 0.4448 \\
0.2 & \bfseries0.9025 & \bfseries0.8973 & \bfseries0.3927 \\
0.4 & 0.8960  & 0.8896 & 0.4035 \\
0.6 & 0.8895 & 0.8778 & 0.4148 \\
0.8 & 0.8789 & 0.8750  & 0.4334 \\
1.0 & 0.8703 & 0.8646  & 0.4470 \\
\bottomrule
\end{tabular}
\end{table}

\section{CONCLUSION}
\label{sec:conclusion}

In this paper, we propose a deep neural network based BIQA model for the surveillance images, which considers the visual saliency characteristics. As the distortions in the salient regions have a great impact on the overall perceptual quality, we first select the local region with the maximum saliency weight via the saliency map. Then, the CNN model and FC network are utilized to predict the global quality for the whole image and local quality for the local region. Finally, the overall quality score is computed as the weighted sum of the global and local quality scores. Experimental results show that the proposed method is effective at predicting the visual quality of the SIs. The proposed BIQA model can help to filter the low-quality SIs and improve the performance of the intelligent video system.

\newpage
%
%
%
\bibliographystyle{splncs04_unsort}
\bibliography{refs}
%




\end{document}